\documentclass[12pt]{amsart}
\usepackage{amssymb,latexsym}

\begin{document}
\title[Wormholes supported by small amounts of exotic matter]
{Wormholes supported by small amounts of exotic matter: some corrections}
\author{Peter K. F. Kuhfittig}
\address{Department of Mathematics\\
Milwaukee School of Engineering\\
Milwaukee, Wisconsin 53202-3109}
\date{\today}

\begin{abstract}
It was pointed out by Fewster and Roman that some of the wormhole models
discussed by Kuhfittig suffer from the failure to distinguish proper from
coordinate distances.  One of the advantages of ``designer wormholes" is
that models can be altered.  The purpose of this note is to show that by
adjusting the metric coefficients, some of these problems can be 
corrected.  By doing so, the basic idea can be retained: wormholes 
containing only small amounts of exotic matter can still be traversable.
\end{abstract}

\maketitle 

PAC number(s): 04.20.Jb, 04.20.Gz

\phantom{a}

In a recent paper Fewster and Roman \cite{FR05} discuss wormholes
models by Visser \cite{VKD03} and Kuhfittig \cite{pK99, pK02, pK03} 
with small amounts, perhaps even arbitrarily small amounts, of exotic 
matter.  It was found that two of the Kuhfittig models are seriously 
flawed by not taking into account the proper distances in estimating 
the size of certain wormholes.  In this note we modify the model in 
Ref. \cite{pK03} in order to salvage the original claim: a wormhole 
containing only small amounts of exotic matter can be traversable 
for humanoid travelers.

Our starting point is the line element \cite{pK03}
\begin{equation}\label{E:line}
   ds^2 =-e^{2\gamma(r)}dt^2+e^{2\alpha(r)}dr^2+r^2(d\theta^2+
      \text{sin}^2\theta\, d\phi^2),
\end{equation}
where $\gamma(r)$ is the redshift function, which must be everywhere
finite.  The function $\alpha$ has a vertical asymptote at $r=r_0$:
$\lim_{r \to r_0+}\alpha(r)=+\infty$.  The shape function can be written 
\[
    b(r)=r(1-e^{-2\alpha(r)}).
\]
In the absence of any other conditions the proper radial distance $l(r)$
to the throat from any point outside, given by
\[
   l(r)=\int\nolimits_{r_0}^{r}e^{\alpha(r')}dr',
\]
may diverge as $r\rightarrow r_0$.  One way to avoid this problem is 
to start with the function
\[
  \alpha_1(r)=\text{ln}\frac{K}{(r-r_0)^a},\qquad 0<a<1,
\]
in the vicinity of the throat; $K$ is a constant having the same units
as $(r-r_0)^a$ and will be determined later.  Away from the throat this
function will have to be joined smoothly to some $\alpha_2(r)>0$ which
goes to zero as $r\rightarrow \infty$.  (This construction will be made 
explicit below.) 

Now observe that for $\alpha_1(r)$
\begin{equation}\label{E:proper}
   l(r)=\int\nolimits_{r_0}^{r}e^{\text{ln}\frac{K}{(r'-r_0)^a}}dr'
    =\int\nolimits_{r_0}^{r} \frac{K}{(r'-r_0)^a}dr',
\end{equation}
which is finite for $0<a<1$.  In fact, $l(r_0)=0$.  The redshift function
is assumed to have a similar form in the vicinity of the throat:
\[
   \gamma_1(r)=-\text{ln}\frac{L}{(r-r_2)^b},\qquad 0<b<1,
\]
where $0<r_2<r_0$ to avoid an event horizon at the throat.  

Using the notation in Ref.~ \cite{pK03}, the violation of the weak energy 
condition (WEC) is given by $\rho-\tau<0$, where
\begin{equation}\label{E:WEC}
  \rho-\tau=\frac{1}{8\pi}\left[\frac{2}{r}e^{-2\alpha(r)}
    \left[\alpha'(r)+\gamma'(r)\right]\right].
\end{equation}
(Sufficiently close to the asymptote, $\alpha'(r)+\gamma'(r)$ is 
clearly negative.)  To satisfy the Ford-Roman constraints, we would 
like the WEC to be satisfied outside some small interval $[r_0,r_1]$.  
To accomplish this, choose $r_1$ and construct $\alpha$ and $\gamma$ 
so that
\[
   |\alpha'_1(r_1)|=|\gamma'_1(r_1)|.
\]
It will be shown presently that if 
\[
    b=\frac{r_1-r_2}{r_1-r_0}a,
\]
then $|\alpha'_1(r)|>|\gamma'_1(r)|$ for $r_0<r<r_1$, and 
$|\alpha'_1(r)|<|\gamma'_1(r)|$ for $r>r_1$.  More precisely,
\[
   \alpha'_1(r)=\frac{-a}{r-r_0}<\frac{-a(r_1-r_2)}{r_1-r_0}
    \frac{1}{r-r_2}=-\gamma'_1(r)
\]
for $r_0<r<r_1$; if $r>r_1$, the sense of the inequality is reversed.  
To see this, observe that equality holds for $r=r_1$.  Now multiply 
both sides by $r-r_2$ and denote the resulting left side by $f(r)$, i.e.,
\[
    f(r)=\frac{-a(r-r_2)}{r-r_0}.
\]
 Since $f'(r)>0$, $f(r)$ is strictly increasing, enough to establish the
inequalities.

Given these forms, we learn immediately that $a>\frac{1}{2}$ by checking
the radial tidal constraint (Morris and Thorne (MT) \cite{MT88}) in
the vicinity of the throat, where the most severe constraints tend to 
occur.  Using the forms given in Ref. \cite{pK03},
\begin{multline}\label{E:radial}
  |R_{\hat{r}\hat{t}\hat{r}\hat{t}}|=\left|e^{-2\alpha_1(r)}
   \left(\gamma''_1(r)-\alpha'_1(r)\gamma'_1(r)
      +\left(\gamma'_1(r)\right)^2\right)\right|\\
    =\left|e^{-2\,\text{ln}K/(r-r_0)^a}\left(\frac{-b}{(r-r_2)^2}
       -\frac{-a}{r-r_0}\frac{b}{r-r_2}+\frac{b^2}{(r-r_2)^2}\right)
            \right|\\
  =\left|\frac{(r-r_0)^{2a}}{K^2}\left(-\frac{b}{(r-r_2)^2}+\frac{ab}
     {(r-r_0)(r-r_2)}+\frac{b^2}{(r-r_2)^2}\right)\right|. 
\end{multline}
As explained in Ref.~\cite{MT88}, we must have  
$|R_{\hat{r}\hat{t}\hat{r}\hat{t}}|\le (10^8 \text{m})^{-2}$.  As long
as $a>\frac{1}{2}$, $(r-r_0)^{2a}/(r-r_0)$ and hence 
$|R_{\hat{r}\hat{t}\hat{r}\hat{t}}|$ is close to zero near the throat.
(The constraint is also met away from the throat, as will be checked 
below.)  Following MT~\cite{MT88}, the wormhole must meet some 
additional conditions: the space stations must be far enough away from 
the throat so that $b(r)/r\approx 0$, making the space nearly flat.  A 
similar formulation is that $1-b(r)/r$ is close to unity:
\begin{equation}\label{E:flat}
   1-\frac{b(r)}{r}=e^{-2\alpha(r)}\approx 0.99.
\end{equation}
In addition, the gradient of the redshift function must meet the 
constraint
\begin{equation}\label{E:gradient}
   |\gamma'(r)|\le g_{\oplus}/\left(c^2\sqrt{1-b(r)/r}\right)=
       1.09\times 10^{-16}\,\text{m}^{-1}
\end{equation}
at the stations \cite{MT88}.

Unfortunately, the above $\alpha_1$ and $\gamma_1$ will eventually 
become negative.  Anticipating this, we can cut these functions off at
some $r=r_3$ and then connect them smoothly to suitable new functions
$\alpha_2$ and $\gamma_2$.
(For physical reasons $r_3$ should be larger than $r_1$.)  Suppose 
$\alpha_2$ has the following form:
\[
    \alpha_2(r)=\frac{A}{r-r_0};
\]
then
\[
    \alpha'_2(r)=-\frac{A}{(r-r_0)^2}.
\]
Since we want $\alpha'_1(r_3)=\alpha_2'(r_3)$, we have
\[
   \alpha'_2(r_3)=-\frac{A}{(r_3-r_0)^2}=\frac{-a}{r_3-r_0}.
\]
 It follows that $A=a(r_3-r_0)$ and
\[
      \alpha_2(r)=\frac{a(r_3-r_0)}{r-r_0}.
\]
Similarly,
\[
    \gamma_2(r)=-\frac{b(r_3-r_0)}{r-r_0}.
\]
Besides having slopes of equal absolute value at $r_3$, we want the
functions to meet, i.e., we want
\[
  \alpha_1(r_3)=\alpha_2(r_3)\,\,\text{and}\,\,\gamma_1(r_3)
       =\gamma_2(r_3).
\]
To this end we must determine $K$ and $L$.  For the first case,
\[
    e^{\text{ln}\frac{K}{(r_3-r_0)^a}}=e^{\frac{a(r_3-r_0)}{r_3-r_0}}
     =e^a;
\]
thus $K=e^a(r_3-r_0)^a$.  In similar manner, $L=e^b(r_3-r_0)^b$.

We are now in a position to estimate the size of the wormhole and to 
check the constraints.  As 
we have seen, $a>\frac{1}{2}$ and $b>a$.  To ensure that $b$ remains 
less than unity, we choose $r_2$ close
enough to $r_0$ so that
\[
    b=\frac{r_1-r_2}{r_1-r_0}a
\]
does not exceed unity.  For the purpose of estimating the distances, 
however, it is convenient to let $a=b=1/2$.  We also assume that 
$r_0\approx 0$: since we are using only small amounts of exotic
matter, $r_0$ is likely to be small compared to the overall size
of the wormhole. 

Simple trial and error show that a good choice for $r_3$ is 
$r_3=0.00005$ l.y.  Then, by Eq. (\ref{E:flat})
\[
  1-\frac{b(r)}{r}=e^{-2\alpha(r)}=e^{-\frac{2ar_3}{r}}
     =e^{-2(1/2)(0.00005)/r}=0.99.
\]
Solving for $r$, we get $r=0.00497$ l.y. for the coordinate distance 
to the space station.  (For comparison, if $a=b=0.75$, we obtain the
slightly larger value $r=0.00746$ l.y.)

The other constraint [Eq.~(\ref{E:gradient})] becomes
\begin{equation*}
   |\gamma'(r)|=\frac{br_3}{r^2}=\frac{\frac{1}{2}(0.00005)
    \times 9.46\times 10^{15}}{(0.00497\times9.46\times 10^{15})^2}
    =1.07\times 10^{-16} \text{m}^{-1}.
\end{equation*}

Returning briefly to Eq.~(\ref{E:radial}), on the interval 
$[r_0,r_3]$, the value of $|R_{\hat{r}\hat{t}\hat{r}\hat{t}}|$ 
becomes zero at the throat, as we have seen, but it does attain 
a maximum value near $r_0$. For the given value of $K$ and hence 
of $r_3$ the constraint can be met there if $a$ and $r_2$ are 
chosen properly. 

For larger values the constraint is more easily met, as one would 
expect.  For example, if 
$r \ge r_3$, then, using $\alpha_2$ and $\gamma_2$, 
\begin{equation*}
   |R_{\hat{r}\hat{t}\hat{r}\hat{t}}|=\left|e^{-2\alpha_2(r)}
   \left(\gamma''_2(r)-\alpha'_2(r)\gamma'_2(r)+(\gamma'_2(r))^2
   \right)\right|.
\end{equation*}
The smallest permissible value is $r=r_3$.  So if $a=b=1/2$, then
\[
    |R_{\hat{r}\hat{t}\hat{r}\hat{t}}|=(2e)^{-1}(r_3\times 9.46\times 
    10^{15}\,\text{m})^{-2}\approx 8.2\times10^{-25}\,\text{m}^{-2}.
\]

Next, we compute the proper radial distance from $r_3$ to the stations,
using $\alpha_2(r)$:
\[
   \int\nolimits_{0.00005}^{0.00497}e^{(1/2)(0.00005)/r}dr=0.00504 
      \,\,\text{l.y.}
\]
To determine the proper radial distance from the throat to $r_3$, we 
use $\alpha_1(r)$:
\begin{multline*}
  \int\nolimits_{r_0}^{r_3}e^{\text{ln}\frac{K}{(r-r_0)^a}}dr
   =\int\nolimits_{r_0}^{r_3}\frac{K}{(r-r_0)^a}dr
    =K\frac{1}{1-a}(r_3-r_0)^{1-a}\\
       =\frac{1}{1-a}e^a(r_3-r_0).
\end{multline*}
To estimate this value, let $a=1/2$, $r_0=0$, and $r_3=0.00005$; then
we get 0.00016 l.y. for the proper distance.  The total proper 
distance from the platform to the throat is 0.0052 l.y.  This estimate 
is in line with those in Refs.~\cite{pK02, pK03}.  In fact, if we 
follow the well-known travel scheme in MT~\cite{MT88}, we find that 
the throat can be reached in about 50 days.

Summarizing these results, since $r_1$ can be arbitrarily chosen, it 
seems as though the exotic region can be made arbitrarily small.  
Unfortunately, besides producing wormholes with some rather 
undesirable features \cite{FR96}, a small enough region would result 
in a value of $r_2$ that is too close to $r_0$, resulting, in turn, in a 
large time dilation: in the vicinity of the throat a proper time interval 
$\Delta\tau$ would correspond to a large coordinate time interval $\Delta t$.
Of course, to the left of $r_1$, $\gamma_1(r)$ could simply be replaced 
by a curve that is much less steep near the throat.  It is nevertheless
best to stay with the claim in the opening paragraph: a wormhole 
supported by only small amounts of exotic matter can still be
traversable.

\end{document}